\documentclass[11pt]{article}
\usepackage{latexsym}
\usepackage[pdftex]{graphicx}
\usepackage{amsfonts}
\usepackage{amsmath, amsthm, amssymb}
\usepackage{fullpage}
\usepackage{authblk}
\usepackage{setspace}
\usepackage{longtable}
\usepackage{natbib}
\usepackage{tikz,pgfplots}
\usetikzlibrary{shapes,decorations,arrows,calc,arrows.meta,fit,positioning}
\tikzset{
    -Latex,auto,node distance =1 cm and 1 cm,semithick,
    state/.style ={ellipse, draw, minimum width = 0.7 cm},
    point/.style = {circle, draw, inner sep=0.04cm,fill,node contents={}},
    bidirected/.style={Latex-Latex,dashed},
    el/.style = {inner sep=2pt, align=left, sloped}
}
\bibpunct{(}{)}{;}{a}{}{,}
\usepackage{dcolumn}
\newcolumntype{.}{D{.}{.}{-1}}
\newcolumntype{d}[1]{D{.}{.}{#1}}
\usepackage[top=.8in, left=.9in, right=.9in,
bottom=1in]{geometry}
\usepackage{bm}
\usepackage[compact]{titlesec}
\usepackage{booktabs}

\usepackage{color}
\RequirePackage[colorlinks,citecolor=blue,urlcolor=blue]{hyperref}
\usepackage{amsthm}

\newtheorem{theorem}{Theorem}

\newtheorem{condition}[theorem]{Condition}

\usepackage{caption}
\usepackage{subcaption}

\expandafter\def\expandafter\normalsize\expandafter{\normalsize\setlength\abovedisplayskip{0pt}}
\expandafter\def\expandafter\normalsize\expandafter{\normalsize\setlength\belowdisplayskip{0pt}}
\expandafter\def\expandafter\normalsize\expandafter{\normalsize\setlength\abovedisplayshortskip{0pt}}
\expandafter\def\expandafter\normalsize\expandafter{\normalsize\setlength\abovedisplayshortskip{0pt}}

\begin{document}
\pagestyle{plain}

\title{On Mendelian Randomisation Mixed-Scale Treatment Effect Robust Identification (MR MiSTERI) and Estimation for Causal Inference}
\author[1]{Zhonghua Liu}
\affil[1]{Department of Statistics and Actuarial Science, University of Hong Kong, Hong Kong}
\author[2]{Ting Ye}
\affil[2,5]{Department of Statistics, The Wharton School of Business, University of Pennsylvania, Philadelphia, PA, USA}
\author[3]{Baoluo Sun}
\affil[3]{Department of Statistics and Applied Probability, National University of Singapore, Singapore}
\author[4,6]{Mary Schooling}
\affil[4]{CUNY Graduate School of Public Health and Health Policy, New York, NY, USA }
\affil[6]{School of Public Health, Li Ka Shing Faculty of Medicine,  University of Hong Kong, Hong Kong}
\author[5*]{Eric Tchetgen Tchetgen}
\affil[*]{ett@wharton.upenn.edu}

\maketitle
\thispagestyle{empty}

\abstract{
\noindent
Standard Mendelian randomization analysis can produce biased results if the genetic variant defining instrumental variable (IV) is confounded and/or has a horizontal pleiotropic effect on the outcome of interest not mediated by the treatment.  We provide novel identification conditions for the causal  effect of a treatment in the presence of unmeasured confounding by leveraging  an invalid IV for which both the IV independence and exclusion restriction assumptions may be violated. The proposed Mendelian Randomization Mixed-Scale Treatment Effect Robust Identification (MR MiSTERI) approach relies on (i) an assumption that the treatment effect does not vary with the invalid IV on the additive scale; and (ii) that the selection bias due to confounding does not vary with the invalid IV on the odds ratio scale; and (iii) that the residual variance for the outcome is heteroscedastic and thus varies with the invalid IV. Although assumptions (i) and (ii) have, respectively appeared in the IV literature, assumption (iii) has not; we formally establish that their conjunction can identify a causal effect even with an invalid IV subject to pleiotropy. MR MiSTERI is shown to be particularly advantageous in the presence of pervasive heterogeneity of pleiotropic effects on the additive scale. For estimation, we propose a simple and consistent three-stage estimator that can be  used as preliminary estimator to a carefully constructed one-step-update estimator, which is guaranteed to be more efficient under the assumed model.  In order to incorporate multiple, possibly correlated and weak IVs, a common challenge in MR studies, we develop a MAny Weak Invalid Instruments (MR MaWII MiSTERI) approach for strengthened identification and improved estimation accuracy.   Both simulation studies and UK Biobank data analysis results demonstrate the robustness of  the proposed MR MiSTERI method.  
}\\

\clearpage

\section{Introduction}
Mendelian randomization (MR) is an instrumental variable (IV) approach that uses genetic variants, for example, single-nucleotide polymorphisms (SNPs), to infer the causal effect of a modifiable risk treatment on a health outcome \citep{Davey-Smith:2003aa}. MR has recently gained popularity in epidemiological studies because, under certain conditions, it can provide unbiased estimates of causal effects even in the presence of unmeasured exposure-outcome confounding. For example, findings from a recent MR analysis assessing the causal relationship between low-density lipoprotein cholesterol  and coronary heart disease \citep{Ference:2017aa} in an observational study are consistent with the results of earlier randomized clinical trials \citep{Scandinavian:1994aa}. 

\begin{figure}[t]
	\centering
	\begin{minipage}{0.49\textwidth}
		\centering
		\begin{tikzpicture}
		\node[state] (1) {$Z$};
		\node[state] (2) [right =of 1] {$A$};
		\node[state] (3) [right =of 2] {$Y$};
		\node[state] (5) [above =of 2,xshift=.9cm, yshift=-0.2cm] {$U$};
		\path (1) edge node[above] {} (2);
		\path (2) edge node[above] {} (3);
		\path (5) edge node[el,above] {} (2);
		\path (5) edge node[el,above] {} (3);
		\end{tikzpicture} \\ \vspace{6mm}(a)
	\end{minipage}\hfill
	\begin{minipage}{0.49\textwidth}
		\centering
		\begin{tikzpicture}
		\node[state] (1) {$Z$};
		\node[state] (2) [right =of 1] {$A$};
		\node[state] (3) [right =of 2] {$Y$};
		\node[state] (5) [above =of 2,xshift=.9cm, yshift=-0.2cm] {$U$};
		\path (1) edge node[above] {} (2);
		\path (2) edge node[above] {} (3);
		\path (5) edge node[el,above] {} (2);
		\path (5) edge node[el,above] {} (3);
		\path (5) edge node[el,above] {} (1);
		\path (1) edge[bend right=40] node[el,above] {} (3);
		\end{tikzpicture}\\(b)
	\end{minipage}
	\caption{Directed acyclic graph (DAG) with an instrument ($Z$), an outcome ($Y$), a treatment ($A$) and unmeasured confounders ($U$). The left panel shows a valid Mendelian randomization study and the right panel shows violations of IV independence and exclusion restriction assumptions. }
	\label{fig: dag}
\end{figure}
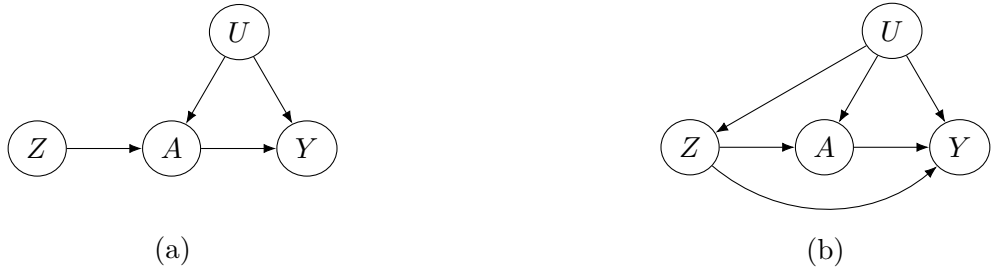

For a SNP to be a valid IV, it must satisfy the following three core assumptions \citep{Didelez:2007aa, Lawlor:2008aa}:
\begin{description}
	\item[(A1)]IV relevance: the SNP must be associated (not necessarily causally) with the exposure;
	\item[(A2)] IV independence: the SNP must be independent of any unmeasured confounder of the exposure-outcome relationship;
	\item[(A3)] Exclusion restriction: the SNP cannot have a direct effect on the outcome variable not mediated by the treatment, that is, no horizontal pleiotropic effect can be present.
\end{description}
The causal diagram in Figure 1(a) graphically represents the three core assumptions for a valid IV. It is well-known that even when assumptions (A1)-(A3) hold for a given IV, the average causal effect of the treatment on the outcome cannot be point identified without an additional condition, the nature of which dictates the interpretation of the identified causal effect. Specifically, \cite{Angrist:1996aa} proved that under (A1)-(A3) and a monotonicity assumption about the IV-treatment relationship, the so-called complier average treatment effect is nonparametrically identified. More recently, \cite{WangETT:2018aa} established identification of population average causal effect under (A1)-(A3) and an additional assumption of no unmeasured common effect modifier of the association between the IV and the endogenous variable, and the treatment causal effect on the outcome. A special case of this assumption is that the association between the IV and the treatment variable is constant on the additive scale across values of the unmeasured confounder (\cite{Tchetgen2020_GENIUS}).  In a separate strand of work, \cite{Robins:1994aa} identified the effect of treatment on the treated under (A1)-(A3)  and a so-called “no current-treatment value interaction” assumption  (A.4a) that the effect of treatment on the treated is constant on the additive scale across values of the IV. In contrast, \cite{Liu_2020} established identification of the effect of treatment on the treated (ETT) under (A1)-(A3), and an assumption (A.4b) that the selection bias function defined as the odds ratio association between the potential outcome under no treatment and the treatment variable, is constant across values of the IV. Note that under the IV DAG Figure \ref{fig: dag}(a), assumption (A1) is empirically testable while (A2) and (A3) cannot be refuted empirically without a different assumption being imposed \citep{Glymour:2012aa}. Possible violation or near violation of assumption (A1) known as the weak IV problem poses an important challenge in MR studies as the associations between a single SNP IV and complex traits can be fairly weak \citep{Staiger:1997aa,Stock:2002aa}. But massive genotyped datasets have provided many such weak IVs. This has motivated a rich body of work addressing the weak IV problem under a many weak instruments framework, from a generalized method of moments perspective given individual level data \citep{Chao:2005aa, Newey:2009aa, Davies:2015aa}, and also from a summary-data perspective \citep{Zhao:2019aa, zhao2018statistical, Ye:2019}. Violation of assumption (A2) can occur due to  population stratification, or when a selected SNP IV is in linkage disequilibrium (LD) with another genetic variant which has a direct effect on the outcome \citep{Didelez:2007aa}. Violations of  (A3) can occur when the chosen SNP IV has a non-null direct effect on the outcome not mediated by the exposure, commonly referred to as horizontal pleiotropy and is found to be widespread \citep{Solovieff:2013aa, Verbanck:2018aa}. A standard MR analysis (i.e. based on standard IV methods such as 2SLS) with an invalid IV that violates any of those three core assumptions might yield biased causal effect estimates. 

Methods to address possible violations of (A2) or (A3) given a single candidate IV are limited. Two methods have recently emerged as potentially robust against such violation under certain conditions. The first, known as MR-GxE, assumes that one has observed an environmental factor (E), which interacts with the invalid IV in its additive effects on the treatment of interest, and that such interaction is both independent of any unmeasured confounder of the exposure-outcome relationship, and does not operate on the outcome in view \citep{Spiller:2019aa, Spiller2020}. In other words, MR-GxE essentially assumes that while the candidate SNP (the G variable) may not be a valid IV, its additive interaction with an observed covariate constitutes a valid IV which satisfies (A1)-(A3).  In contrast, MR GENIUS relies on an assumption that the residual variance of the first stage regression of the treatment on the candidate IV is heteroscedastic with respect to the candidate IV, i.e., the variance of the treatment depends on the IV, an assumption that may be viewed as strengthening of the IV relevance assumption \citep{Lewbel2012,Tchetgen2020_GENIUS}. Interestingly, as noted by \cite{Tchetgen2020_GENIUS}, existence of a GxE interaction that satisfies conditions (A1)-(A3) required by MR GxE had such an E variable (independent of $G$) been observed, necessarily implies that the heteroscedasticity condition required by MR GENIUS holds even when the relevant E variable is not directly observed. Furthermore, as it is logically possible for heteroscedasticity of the variance of the  treatment to operate even in absence of any GxE interaction, MR GENIUS can be valid in certain settings where MR GxE is not. 

In this paper, we develop an alternative robust MR approach for estimating a causal effect of a treatment subject to unmeasured confounding, leveraging a potentially invalid IV which fails to fulfill either IV independence or exclusion restriction assumptions. The proposed “Mendelian Randomization Mixed-Scale Treatment Effect Robust Identification” (MR MiSTERI) approach relies on (i) an assumption that the treatment effect does not vary with the invalid IV on the additive scale; and (ii) that the selection bias due to confounding does not vary with the invalid IV on the odds ratio scale; and (iii) that the residual variance for the outcome is heteroscedastic and thus varies with the invalid IV. Note that  assumption (iii) is empirically testable. Although assumptions (i) and (ii) have, respectively appeared in the IV literature, assumption (iii) has not; we formally establish that their conjunction can identify a causal effect even with an invalid IV subject to pleiotropy. MR MiSTERI is shown to be particularly advantageous in the presence of pervasive heterogeneity of pleiotropic effects on the additive scale, a setting in which both MR GxE and MR GENIUS can be severely biased whenever heteroscedastic first stage residuals can fully be attributed to latent heterogeneity in SNP-treatment association \citep{Spiller2020}. For estimation, we propose a simple and consistent three-stage estimator that can be  used as a preliminary estimator to a carefully constructed one-step-update estimator, which is guaranteed to be more efficient under the assumed model.  In order to incorporate multiple, possibly correlated and weak IVs, a common challenge in MR studies, we develop a MAny Weak Invalid Instruments (MaWII MR MiSTERI) approach for strengthened identification and improved accuracy.  Simulation study results show that our proposed MR method gives consistent estimates of the causal parameter and the selection bias parameter with nominal confidence interval coverage with an invalid IV, and the accuracy is further improved with multiple weak invalid IVs. For illustration, we apply our method to the UK Biobank data set to estimate the causal effect of body mass index (BMI) on average glucose level. We also develop an R package freely available for public use at \url{https://github.com/zhonghualiu/MRMiSTERI}.

\section{Methods} 
\subsection{Identification with a Binary Treatment and a Possibly Invalid Binary IV } 
Suppose that we observe data $O_i=(Z_i, A_i, Y_i)$ of sample size $n$  drawn independently from a common population, where $Z_i, A_i, Y_i$  denote a SNP IV, a treatment and a continuous outcome of interest for the $i$th subject ($1\leq i\leq n$), respectively. In order to simplify the presentation, we drop the sample index $i$   and do not consider observed covariates at this stage, although all the conclusions continue to hold within strata defined by observed covariates. Let $z, a, y$  denote the possible values that $Z, A, Y$  could take, respectively. Let $Y_{az}$  denote the potential outcome, had possibly contrary to fact,  $A$  and $Z$ been set to $a$  and  $z$ respectively, and let $Y_a$  denote the potential outcome had $A$  been set to $a$. We are interested in estimating the ETT defined as 
\[
\beta(a)= E(Y_a- Y_0\mid A=a).
\]

To facilitate the exposition, consider the simple setting where both the treatment and the SNP IV are binary, then the ETT simplifies to  $\beta= E(Y_1- Y_0\mid A=1)$. By the consistency assumption \citep{Hernan-Robins},  we know that  $E(Y_1\mid A=1)=E(Y\mid A=1)$. However, the expectation of the potential outcome $Y_0$  among the exposed subpopulation $E(Y_0\mid A=1)$ cannot empirically be observed   due to possible unmeasured confounding for the exposure-outcome relationship. The following difference captures this confounding bias on the additive scale ${Bias}=E(Y_0\mid A=1)- E(Y_0\mid A=0)$, which is exactly zero only when exposed and unexposed groups are exchangeable on average (i.e. under no confounding) \citep{Hernan-Robins}, and is otherwise not null. With this representation and the consistency assumption, we have 
\[
E(Y\mid A=1)- E(Y\mid A=0)=\beta+ Bias.
\]

This simple equation implies that one can only estimate the sum of the causal effect $\beta$  and the confounding bias but cannot tease them apart using the data $(A, Y)$ only. With the availability of a binary candidate SNP IV $Z$ that is possibly invalid, we can further stratify the population by $Z$ to obtain under consistency:
\[
E(Y\mid A=1, Z=z)-  E(Y\mid A=0, Z=z)=\beta(z)+Bias(z),
\]
where $z$ is equal to either 0 or 1, and $\beta(z)= E(Y_a- Y_0\mid A=a, Z=z)$, $Bias(z)=E(Y_0\mid A=1, Z=z)- E(Y_0\mid A=0, Z=z)$ denote the causal effect and bias in the stratified population with the IV taking value $z$. Note that there are only two equations but four unknown parameters:  $\beta(z)$,  $Bias(z), z=0,1$. Therefore, the causal effect cannot be identified without imposing assumptions to reduce the total number of parameters to two. 

Our first assumption extends the no current-treatment value interaction assumption (A.4a) originally proposed by \cite{Robins:1994aa}, that the causal effect does not vary across the levels of the SNP IV so that $\beta(z)$  is a constant as function of $z$. Formally stated: 
\begin{description}
	\item[(B1)]Homogeneous ETT assumption: $E(Y_{a=1, z}- Y_{a=0, z}\mid A=1, Z=z)=\beta$.
\end{description}
It is important to note that this assumption does not imply the exclusion restriction assumption (A3); it is perfectly compatible with presence of a  direct effect of $Z$ on $Y$ (the direct arrow from $Z$ to $Y$ is present in Figure \ref{fig: dag}(b)), i.e. $E(Y_{a=0, z=1}- Y_{a=0, z=0}\mid A=1, Z=z)\neq 0$ as well as unmeasured confounding of the effects of Z on (A,Y), both of which we accommodate.

In order to state our second core assumption, consider the following data generating mechanism for the treatment, which encodes presence of unmeasured confounding by making dependence between $A$ and $Y_0$ explicit under a logistic model:  
\[
\text{logit}\left\{Pr(A=1\mid Y_0=y_0, Z=z)\right\}= \gamma_0+\gamma y_0+\gamma_z z +\gamma_{y_0z}y_0 z.
\]

\begin{figure}
	\centering
	\includegraphics[scale=.6]{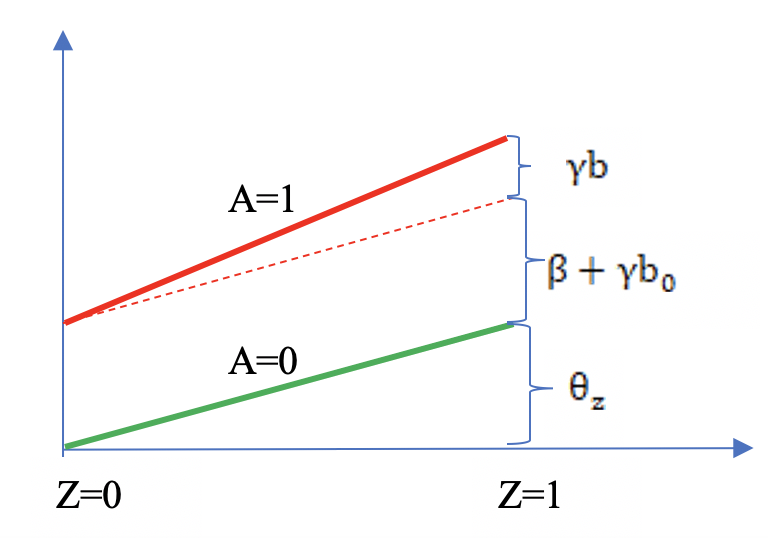}
	\caption{An illustration of our identification strategy for a binary treatment $A$ and binary IV $Z$. The green line represents  $E(Y\mid A=0, Z=z)$ and its difference between $Z=1$ and $Z=0$ is  $\theta_z$. The red dashed line (parallel to the green line) represents the hypothetical setting when $\sigma^2(Z=z)$  is constant, and the red solid line represents the setting when  $\sigma^2(Z=z)$  is not a constant function in $Z$. Note that $b_0$  and  $b$ are treated as known. }
	\label{fig:identification}
\end{figure}

This model can of course not be estimated directly from the observed data without any additional assumption because it would require the potential outcome  $Y_0$ be observed both on the untreated (guaranteed by consistency) and the treated. Nevertheless, this model implies that the conditional (on $Z$) association between $A$ and $Y_0$  on the odds ratio scale is $OR(Y_0=y_0, A=1\mid Z=z)=\exp(\gamma y_0+ \gamma_{y_0 z}y_0 z)$. Together, $\gamma$  and  $\gamma_{y_0 z}$ encode the selection bias due to unmeasured confounding on the log-odds ratio scale. If both $\gamma$  and  $\gamma_{y_0 z}$ are zeros, then $A$  and  $Y_0$ are conditionally (on $Z$) independent, or equivalently, there is no residual confounding bias upon conditioning on $Z$. Our second identifying assumption formally encodes assumption A4.b of \cite{Liu_2020} of homogeneous odds ratio selection bias   $\gamma_{y_0 z}=0$: 
\begin{description}
	\item[(B2)]Homogeneous selection bias:  $OR(Y_0=y_0, A=a\mid Z=z)=\exp(\gamma a y_0)$.
\end{description}
Assumption (B2) states that the selection bias on the odds ratio scale is homogeneous across levels of the IV. Thus, this assumption allows for the presence of unmeasured confounding which, upon setting $\gamma_{y_0 z}=0$ is assumed to be on average balanced with respect to the SNP IV (on the odds ratio scale).  Following \cite{Liu_2020}, we have that under (B2) 
\[
E(Y_0\mid A=a, Z=z)= \frac{E\{ Y\exp(\gamma aY)\mid A=0, Z=z\}}{E\{\exp(\gamma aY)\mid A=0, Z=z\}}
\]
Define $\varepsilon= Y- E(Y\mid A, Z)$, and suppose that $\varepsilon\mid A, Z\sim N(0, \sigma^2( Z))$, then after some algebra we have,
\[
E(Y_0\mid A=1, Z=z)= E(Y\mid A=0, Z=z)+\gamma\sigma^2(Z).
\]
The selection bias on the odds ratio scale does not vary with the levels of the IV, however in order to achieve identification, the bias term  $\gamma \sigma^2( Z)$  must depend on the IV. This observation motivates our third assumption,
\begin{description}
	\item[(B3)] Outcome heteroscedasticity, that is $\sigma^2( Z)$ cannot be a constant.
\end{description}
Assumption (B3) is empirically testable using existing statistical methods for detecting non-constant variance in regression models.  As shown by \cite{Pare2010}, observed or unobserved  gene–gene (GxG) and/or gene–environment (GxE) can result in changes in variance of a quantitative trait per genotype.  \cite{Pare2010} found several SNPs associated with the variance of inflammation markers.  Therefore,   GxE interaction can be inferred from genetic variants associated with phenotypic variability without the need of measuring environmental factors, usually termed variance quatitative trait loci (vQTL) screening \citep{Marderstein2021}. In particular,  \cite{Wang2019}  performed a genome-wide vQTL analysis of  about 5.6 million variants on 348,501 unrelated individuals of European ancestry in the UK Biobank and identified 75 significant vQTLs for 9   quantitative traits (out of 13 traits under study). 
Hence, one can generally expect this novel identification assumption (B3) to hold in the presence of pervasive heterogeneity of pleiotropic effects on the additive scale for the outcome, a setting in which both MR GxE and MR GENIUS can be severely biased.   
Under assumptions (B1)-(B3), we have 
\[
E(Y\mid A=1, Z=z)- E(Y\mid A=0, Z=z) =\beta+ \gamma\sigma^2( z).
\]
Note that  $\sigma^2( z)$ is the variance of Y given z, and thus can be estimated with no bias. For the binary IV $Z$,  $\sigma^2(Z=0)$ and  $\sigma^2( Z=1)$ are the variance of $Y$ within each IV stratum and can easily be estimated using the sample variance in each group. We will describe estimating procedures in the next section for more general settings where $Z$ is not necessarily binary. Importantly, in the present case, we now have two equations and two unknown parameters $\beta, \gamma$. 
Denote $D(Z=z)= E(Y\mid A=1, Z=z)- E(Y\mid A=0, Z=z)$. We have the following  main identification result. 
\begin{theorem}
	Under assumptions (B1)-(B3), the selection bias parameter and the causal effect of interest are uniquely identified as follows:
	\begin{align*}
	\gamma&= \frac{D(Z=1)- D(Z=0)}{\sigma^2( Z=1)-\sigma^2( Z=0) } \\
	\beta&=E \{ D(Z)- \frac{D(Z=1)- D(Z=0)}{\sigma^2( Z=1)-\sigma^2( Z=0) } \sigma^2( Z)\}\\
	&= D(z)- \frac{D(Z=1)- D(Z=0)}{\sigma^2( Z=1)-\sigma^2( Z=0) } \sigma^2( z); z=0,1.
	\end{align*}
\end{theorem}
In an observed sample, one can simply use the sample versions of unknown quantities to obtain estimates for $\beta$  and $\gamma$ respectively.  Standard errors can be deduced by a direct application of the multivariate delta method, or by using resampling techniques such as the jackknife or the bootstrap. \\
{\bf{Remark 1.}} Note that as stated in the theorem, the causal parameter $\beta$  can be estimated based on data among either $Z=0$ or $Z=1$ group. In practical settings, in order to improve efficiency, one may take either unweighted average of the two estimates as given in the theorem, or their inverse variance weighted average. In fact, one may fit a saturated linear model for the variance $\sigma^2( Z=z)=b_0+b z$; We give a graphical illustration of our identification strategy as shown in Figure \ref{fig:identification}. By using the outcome heteroscedasticity assumption (B3), we can identify the selection bias parameter $\gamma$  and then the causal effect parameter $\beta$. Without the assumption (B3) as shown by the red dashed line, we can only obtain  $\beta+\gamma b_0$ and thus cannot tease apart the causal effect  and the selection bias. 

\subsection{Identification with a Continuous Treatment and a Possibly Invalid Discrete IV }
The identification strategy in the previous section extends to the more general setting of a continuous treatment and a more general discrete IV, for example, the SNP IV taking values 0, 1, 2 corresponding to the number of minor alleles. To do so,  we introduce the notion of generalized conditional odds ratio function as a measure of the conditional association between a continuous treatment and a continuous outcome. To ground idea, consider the following model for the treatment free potential outcome: 
\begin{align*}
Y_0=E(Y_0|A,Z)+\sigma(Z)\epsilon
\end{align*}
where $\epsilon$ is independent standard normal; then one can show that the generalized conditional odds ratio function associated with $Y_0$ and $A$ given $Z$  with $(Y_0=0, A=0)$ taken as reference values \citep{Chen2007:aa} is given by
\[
OR(y_0, a\mid z)= \frac{f(y_0\mid a, z)f(y_0=0\mid a=0, z) }{f(y_0\mid a=0, z)f(y_0=0\mid a, z)}
= \exp\{(E(Y_0|A=a,z)-E(Y_0|A=0,z)) y_0/ \sigma^2(z)\}.
\]
By Assumption (B2), we have   $$E(Y_0|A,Z)-E(Y_0|A=0,Z)=\gamma \sigma^2(Z) A.$$ Therefore, 
\[
\mu(a, z)=E(Y\mid A=a, Z=z)=\beta a +\gamma\sigma^2(Z=z) a+ E(Y\mid A=0, Z=z).
\]
For practical interpretation, we assume $A$ has been centered so that $A=0$ actually represents average treatment value in the study population. Assume the following model $E(Y\mid A=0, Z=z)=\theta_0 +\theta_z z$ and a log linear model for $\sigma^2(Z)$ such that 
\begin{align*}
\mu(a, z)&= \beta A+ \gamma A\sigma^2(z)+\theta_0 +\theta_z z,\\
log(\sigma^2(z))&= \eta_0+\eta_z z.
\end{align*}
The conditional distribution of $Y$ given $A$ and $Z$ is 
\[
Y\mid A=a, Z=z\sim N\left(\mu (a, z), \sigma^2(z)\right).
\]
Then  we can simply use standard maximum likelihood estimation (MLE)  to obtain consistent and fully efficient estimates for all parameters under the assumed model. However, the likelihood may not be log-concave and  direct maximization of the likelihood function with respect to all parameters  might not always converge. Instead, we propose a novel three-stage estimation procedure which can provide a consistent but inefficient preliminary estimator of unknown parameters. We then propose a carefully constructed one-step update estimator to obtain a consistent and fully efficient estimator.  Our three-stage estimation method works as follows. 
\begin{description}
	\item[Stage 1:]Fit the following linear regression using standard (weighted) least-squares  
	\[
	Y= \theta_0+\theta_a A +\theta_z Z+\theta_{az} AZ +\varepsilon.
	\]
	We obtain parameter estimates $\hat \theta_0, \hat\theta_a, \hat\theta_z, \hat\theta_{az}$, $\hat E(Y\mid A=0, Z)= \hat\theta_0+\hat\theta_z Z$, with corresponding estimated residuals $\hat\varepsilon= Y-  \hat\theta_0-\hat\theta_a A -\hat
	\theta_z Z-\hat\theta_{az} AZ$. 
	\item[Stage 2:] Regress the squared residuals $\hat\varepsilon^2$  on $(1, Z)$ in a log-linear model and obtain parameter estimates  $\hat\eta_0, \hat\eta_z$  and $\hat\sigma^2(Z)=\exp(\hat\eta_0+ \hat\eta_z Z)$.
	\item[Stage 3:] Regress $Y-\hat E(Y\mid A=0, Z) $  on $A$ and $A\hat\sigma^2(Z)$  without an intercept term and obtain the two corresponding regression coefficients estimates $\hat\beta$  and  $\hat\gamma$ respectively. Alternatively, we may also fit $E(Y\mid A, Z)= \beta A+\gamma A\hat\sigma^2(Z)+ \theta_0+\theta_z Z $.
\end{description}

The proposed three-stage estimation procedure is computationally convenient, appears to always converge and provides a consistent estimator for all parameters under the assumed model. Next we propose the following one-step update to obtain a fully efficient estimator.  Consider the following normal model 
\begin{equation}
\frac{Y- \{ \beta A+ \gamma A \exp(\eta_0+\eta_z Z) + \theta_0 +\theta_z Z\}}{\sqrt{ \exp(\eta_0+\eta_z Z)}} \sim N(0, 1). \label{eq: normal}
\end{equation}
Denote $\Theta= (\beta, \gamma, \eta_0, \eta_z, \theta_0, \theta_z)$ and denote the log-likelihood function as $l(\Theta; Y, A, Z)$  for an individual. With a sample of size $n$, the log-likelihood function is 
\[
l_n(\Theta)= \sum_{i=1}^n l_i (\Theta; Y_i, A_i, Z_i).
\]
The score function $S_n(\Theta)= \partial l_n(\Theta)/\partial\Theta $  is defined as the first order derivative of the log-likelihood function with respect to  $\Theta$, and use $i_n(\Theta)=\partial^2  l_n(\Theta)/\partial \Theta\partial \Theta^T $ to denote the second order derivative of the log-likelihood function. Suppose we have obtained a consistent estimator $\tilde \Theta^{(0)}$   of $\Theta$  using the three-stage estimating procedure, then one can show that a consistent and  fully efficient estimator is given by the one-step update estimator $\tilde\Theta^*$  
\begin{eqnarray}
\tilde{\Theta}^*=\tilde\Theta^{(0)}- \{i_n(\tilde \Theta^{(0)}) \}^{-1} S_n (\tilde \Theta^{(0)})). \label{eq: onestep}
\end{eqnarray}
{\bf Remark 2}. In biomedical studies, continuous outcomes of interest are often approximately normally distributed, and the normal linear regression is also routinely used by applied researchers. When the normal assumption is violated, a typical remedy is to apply a transformation to the outcome prior to modeling it, such as the log-transformation or the more general Box-Cox transformation to achieve approximate normality. Alternatively, in section \ref{sec:mixture} we propose to model the error distribution as a more flexible Gaussian mixture model which is more robust than the Gaussian model. Alternatively, as described in the Supplemental Materials, one may consider a semiparametric location-scale model which allows the distribution of standardized residuals to remain unrestricted, a potential more robust approach although computationally more demanding.

\subsection{Estimation and Inference under Many Weak Invalid IVs}
\label{subsec: weak}
Weak identification bias is a salient issue that needs special attention when using genetic data to strengthen causal inference, as most genetic variants typically have weak association signals. When many genetic variants are available, we recommend using the conditional maximum likelihood estimator (CMLE) based on \eqref{eq: normal}, where $Z$ is replaced with a multi-dimensional vector. Let $ \hat{\Theta} $ be the solution to the corresponding score functions, i.e., $S_n( \hat{\Theta})=0$; let   $$ 
I_1(\Theta)=  -E\left\{ \frac{\partial^2}{\partial \Theta\partial \Theta^T} l_1(\Theta, Y_1, A_1, Z_1)\right\}
$$ 
be the Fisher information matrix based on the conditional likelihood function for one observation. Let $k$ be the total number of parameters in $\Theta$, which is equal to $4+2p$ when $Z$ is $p$ dimensional. When $\lambda_{\min}  \{n I_1(\Theta)\}/k \rightarrow\infty$ with  $ \lambda_{\min} \{n I_1(\Theta)\} $ being the minimum eigenvalue of $ n I_1(\Theta) $, we show in the Supplementary Materials (Theorem 2) that $\hat\Theta$ is consistent and asymptotically normal as the sample size goes to infinity, and satisfies 
\begin{eqnarray}
\sqrt{n} (\hat{\Theta}- \Theta) \xrightarrow{d} N\big(0, \{ I_1 (\Theta)\}^{-1} \big). \label{eq: mle normality}
\end{eqnarray}
In other words, the CMLE is robust to weak identification bias and the usual inference procedure can be directly applied. 

The key condition $\lambda_{\min}  \{n I_1(\Theta)\}/k \rightarrow\infty$ warrants more discussion. Note that the quantity $\lambda_{\min}  \{n I_1(\Theta)\}/k$ measures the ratio between the amount of information that a sample of size $n$ carries about the unknown parameter and the number of parameters. In classical strong identification scenarios where the minimum eigenvalue of $ I_1(\Theta) $ is assumed to be lower bounded by a constant, then the condition simplifies to that the number of parameter is small compared with sample size, i.e., $n/k\rightarrow\infty$. However, when identification is weak, the minimum eigenvalue of $I_1(\Theta)$ can be small. In practice, the condition $\lambda_{\min}  \{n I_1(\Theta)\}/k \rightarrow\infty$ can be evaluated by 
$$\hat\kappa= -\lambda_{\min} \{ i_n (\hat \Theta)\}/k, $$
which is the ratio between  the minimum eigenvalue of the negative Hessian matrix and the number of parameters. We remark that the condition stated in the Supplementary Materials (Theorem 2) is more general and applies to general likelihood-based methods, which, for example, implies the condition for consistency for the profile likelihood estimator (MR.raps) in \cite{zhao2018statistical}. In practice, we recommend checking that the $\hat{\kappa}$ is at least greater than 10 based on our simulation studies. 
For valid statistical inference, a consistent estimator for the  variance covariance matrix for $\hat\Theta$ is simply the negative Hessian matrix $- i_n (\hat\Theta)$, whose corresponding diagonal element estimates the variance of the treatment effect estimator $\hat\beta$. Other variants of the CMLE can also be used, for example, the one-step iteration estimator $\tilde\Theta^*$ in \eqref{eq: onestep} is asymptotically equivalent to the CMLE $\hat\Theta$ as long as the initial estimator $ \tilde\Theta^{(0)}$ is $\sqrt{n}$-consistent \citep{Shao}. 

\subsection{More Robust Gaussian Mixture Model for the Outcome }
\label{sec:mixture}
As mentioned in Remark 2, although many continuous outcomes can be well modelled using normal linear model, however sometimes, a more robust model for the outcome error term is desired. As shown by \cite{Verbeke1996}, a Gaussian mixture model is more general and thus  more robust 
than the normal model for modelling an outcome error distribution. Consider the general location-scale model 
$$Y=E(Y|A,Z)+\sigma(Z)\epsilon,$$ where $\epsilon \perp A,Z$. Under assumptions (B1)-(B3), the conditional mean function is given by
\begin{eqnarray*}
	E\left( Y|A=a,Z=z\right)=\beta a+E\left( Y|A=0,Z=z\right)
	+\frac{\partial}{\partial (\gamma a)}\ln \int \exp \left[ \gamma a\sigma \left(
	Z=z\right) \epsilon \right] dF\left( \epsilon \right),
\end{eqnarray*}%
where $F(\cdot)$ denotes the cumulative distribution function of $\epsilon$. We assume that the outcome error distribution can be reasonably approximated by a Gaussian mixture model with enough components \citep{Verbeke1996}. Specifically, let $f(\epsilon)=\sum_{k=1}^K \pi_k \phi_k(\epsilon)$ where $\phi_k(\cdot)$ is the normal density with mean $\mu_k$ and variance $\delta^2_k$ satisfying the constraints
\begin{eqnarray*}
	\sum_{k=1}^{K}\pi _{k}=1, \quad E(\epsilon)=\sum_{k=1}^{K}\pi _{k}\mu _{k}=0,\quad \text{Var}\left( \epsilon \right) =\sum_{k=1}^K\pi _{k}\left( \delta _{k}^{2}+\mu _{k}^{2}\right) =1.
\end{eqnarray*}
The conditional mean function with Gaussian mixture error is given by
\begin{eqnarray*}
	\beta a+E\left( Y|A=0,Z=z\right)+\frac{\sigma \left(
		Z=z\right) }{\sum_{k=1}^{K}\pi _{k}\omega _{k}}\left\{ \left[
	\sum_{k=1}^{K}\pi _{k}\omega _{k}\mu _{k}\right] \right\} +\gamma
	a\sum_{k=1}^{K}\frac{\pi _{k}\omega _{k}\delta _{k}^{2}\sigma
		^{2}\left( Z=z\right) }{\sum_{k=1}^{K}\pi _{k}\omega _{k}},
\end{eqnarray*}%
where $\omega _{k} =\exp \left( \gamma a \sigma \left( Z=z\right) \mu _{k} +\delta _{k}^{2}\left[ \gamma a \sigma \left( Z=z\right) \right] ^{2}/2\right)$. The conditional distribution is
$\{Y-E\left( Y|A=a,Z=z\right)\}/{\sigma \left(
	Z=z\right) }\sim \sum_{k=1}^{K}\pi _{k}N(\mu _{k},\delta _{k}^{2})$.
In practice, estimation of $(\beta,\gamma)$, which are of primary interest, may proceed under a user-specified integer value $K<\infty$ as well as parametric models for $E(Y|A=0,Z=z)$ and $\sigma(Z=z)$ via an alternating optimization algorithm which we describe in the Supplementary Materials.

\section{Simulation Studies}
In this section, we conduct extensive simulation studies to evaluate the finite-sample performance of MR-MiSTERI compared to its main competitors, standard two stage least squares (TSLS) and MR GENIUS. 
We considered sample sizes of $n= 10000, 30000, 100000$. The IV $Z$ is generated from a Binomial distribution with probability equal to 0.3 (minor allele frequency). The treatment $A$ is generated from a  standard normal distribution  $N(0,1)$. The outcome $Y$ is also generated from a normal distribution with mean and variance compatible with assumptions B1-B3: 
\begin{align*}
E(Y\mid A, Z)&= 0.8 A+ 0.2 A \sigma^2( Z)+1+0.3 Z\\
\sigma^2(Z)&= \exp(0.1+\eta_z Z)
\end{align*}
where $\eta_z$ is set to be 0.2, 0.15, 0.1, 0.05. Under this model, we have that the ETT is equal to $\beta=0.8$ and $\gamma=0.2$. We generated in total 1000 such data sets and applied our proposed method to each of them along with TSLS and MR GENIUS. We found that TSLS gives severely biased estimates of the causal effect with estimates as extreme as $-195.72$ and  standard errors as large as $381.57$. MR GENIUS is also severely biased as the average point estimate for the causal effect is $-14.99$. Numerical results for TSLS and MR GENIUS are expected as their corresponding identification conditions are not met in the simulation setup.  Specifically, both exclusion restriction and heteroscedasticity of the treatment with respect to $Z$  assumptions do not hold.  Simulation results for MR MiSTERI are summarized in Table \ref{table-sim}. We found that when sample size is 10000, the bias and standard error of the causal effect and selection bias parameter estimates become larger when the IV strength  $\eta_z$ decreases from 0.2 to 0.1. The causal effect is less sensitive to the IV strength compared to the selection bias parameter $\gamma$. As we increase sample size, the bias decreases and becomes negligible at n=100000.  Confidence intervals achieved the nominal  95\% coverage.

\begin{table}[]
	\caption{Simulation results for estimating the confounding bias parameter $\gamma$ and the ETT $\beta$ using the one-step estimator approach (without covariates) based on 1000 Monte Carlo experiments with a range of sample sizes n. $\hat{\gamma}$ is the averaged point estimates of $\gamma$, and bias is calculated in percentage format. Likewise for $\hat{\beta}$. SE is the averaged standard error. SD stands for the sample standard deviation of the 1000 point estimates for $\gamma$ and $\beta$. Coverage is calculated as the proportion of 95\% confidence intervals that  contains the true parameter value among those 1000 experiments. We vary the value of $\eta_z$ in $\sigma^2(Z)=exp(\eta_0+\eta_zZ)$) to assess the impact of decreasing  IV strength on the estimation results. }
	\label{table-sim}
	\resizebox{\textwidth}{!}{\begin{tabular}{ccccccccccccc}
			\hline
			n & $\eta_z$&	$(\beta,\gamma)$ & $\hat{\beta}$ &Bias &SE &SD & Cover &$\hat{\gamma}$ &	Bias &	SE &	SD&	Cover\\
			\hline
			1e4 & 0.2  & (0.8,0.2) & 0.806 & 0.74 \% & 0.092 & 0.094 & 93.8\% & 0.196 & -2.12\% & 0.074 & 0.075 & 94.6\% \\
			1e4 & 0.15 & (0.8,0.2) & 0.788 & -1.44\% & 0.124 & 0.126 & 95.6\% & 0.210 & 4.77 \% & 0.103 & 0.104 & 95.6\% \\
			1e4 & 0.1  & (0.8,0.2) & 0.778 & -2.76\% & 0.197 & 0.210 & 96.2\% & 0.219 & 9.32\%  & 0.168 & 0.181 & 96.0\% \\
			3e4 & 0.1  & (0.8,0.2) & 0.789 & -1.35\% & 0.104 & 0.106 & 96.0\% & 0.209 & 4.58 \% & 0.089 & 0.090 & 95.8\% \\
			3e4 & 0.05 & (0.8,0.2) & 0.788 & -1.46\% & 0.226 & 0.222 & 97.0\% & 0.210 & 5.17\%  & 0.199 & 0.195 & 97.4\% \\
			1e5 & 0.05 & (0.8,0.2) & 0.797 & -0.36\% & 0.113 & 0.120 & 95.2\% & 0.203 & 1.29\%  & 0.099 & 0.105 & 95.2\% \\
			\hline
	\end{tabular}}
\end{table}

The second simulation study is designed to evaluate the finite sample performance of the proposed  three-stage estimator and the CMLE in the presence of many weak invalid IVs. The sample size is set to be 100000, each IV $Z_j, j=1,\dots, p$, is still generated to take values 0,1,2 with minor allele frequency equal to 0.3. The treatment $A$ is generated from standard normal distribution.  The outcome $Y$ is generated from a normal distribution with mean and variance under B1-B3: 
\begin{align*}
E(Y\mid A, Z)&= 0.8 A+ 0.2 A \sigma^2(Z)-0.5+0.5\sum_{j=1}^p Z_j,\\
\sigma^2(Z)&= \exp\left(0.1+0.05 \sum_{j=1}^p Z_j\right).
\end{align*}
Simulation results are presented in Table \ref{tb: weak}, where the standard error (SE) for the three-stage estimator is the bootstrap estimator approximated using 100 Monte Carlo simulations, the SE for the CMLE is obtained from the inverse Hessian matrix discussed in Section \ref{subsec: weak}. From results in Table \ref{tb: weak}, the three-stage estimator has negligible bias when $p=20$, but shows evidently larger bias when $p$ grows to 50, which is also reflected by the fact that coverage of $95\%$ CI is substantially lower than its nominal level. In contrast, the CMLE shows negligible bias in both scenarios. The standard errors for the CMLE are close to Monte Carlo standard deviation, and the CMLE estimators show nominal coverage in both simulation scenarios. These observations agree with our theoretical assessment that the CMLE is efficient and is robust to many weak invalid IVs. In particular, it is consistent and asymptotically normal as long as the condition $\hat\kappa$ is reasonably large.  Based on empirical evaluations (in the Supplementary Materials), we recommend checking that the  $\hat\kappa$ value is at least greater than 10. 

We similarly generate  the treatment $A$  from the standard normal distribution, the IV $Z$ which takes values in $\{0,1,2\}$ with minor allele frequency equal to 0.3. However we generate $Y=E(Y|A,Z)+\sigma(Z)\epsilon$ where  $E(Y\mid A, Z) =0.8 A + (1 +0.3 Z) +\sigma(Z)\left\{\frac{\sum_{k=1}^{2}\pi _{k}\omega _{k}\left[ \mu _{k}+\delta_{k}^{2}\gamma A\sigma \left(Z\right) \right] }{\sum_{k=1}^{2}\pi_{k}\omega _{k}}\right\}$, $\sigma^2(Z)= \exp(0.1+\eta_z Z)$, and the error $\epsilon$ is generated from a Gaussian mixture distribution with two components,
\begin{equation*}
\begin{aligned}
\varepsilon_1 &\sim N(\mu_1,\delta^2_1), \quad \varepsilon_2 \sim N(\mu_2,\delta^2_2), \quad K\sim \text{Bernoulli} (p=\pi_1),\quad \epsilon=K \varepsilon_1 +(1-K) \varepsilon_2,
\end{aligned}
\end{equation*}
with the parameter values $(\pi_1,\pi_2)=(0.4,0.6)$, $(\mu_1,\mu_2)=(-0.6,0.4)$ and $(\delta_1,\delta_2)=(0.5,1.049)$. We vary the value of $\eta_z$ in the set $\{0.1,0.25,0.5\}$ to assess the impact of IV strength. The results using the proposed Gaussian mixture method with $K=2$ are summarized in Table \ref{table-sim-mixture}. There is noticeable finite-sample bias and under-coverage when $\eta_z=0.1$, especially for estimation of the selection bias parameter $\gamma$, but the performance improves with larger $\eta_z$ or sample size.

\begin{table}[!h]\centering
	\caption{Simulation results for estimating the confounding bias parameter $\gamma=0.2$ and the ETT $\beta=0.8$ using the three-stage estimator and the CMLE based on 1000 Monte Carlo experiments, with $n=100000$ and varying $p$. The third column is the averaged $\hat\kappa = -\lambda_{\min} \{ i_n (\hat \Theta)\}/k$, which is the consistency and asymptotic normality condition for the CMLE and is preferably to be large. $\hat{\gamma}$ is the averaged point estimates of $\gamma$, and bias is calculated in percentage format. Likewise for $\hat{\beta}$. SE is the averaged standard error. SD stands for the sample standard deviation of the 1000 point estimates for $\gamma$ and $\beta$. Coverage is calculated as the proportion of 95\% confidence intervals that  contains the true parameter value among those 1000 experiments.\label{tb: weak}}
	\resizebox{\textwidth}{!}{\begin{tabular}{cccccccccccccc} \hline
			$n$        & $p$        & $\hat\kappa$ & Method& $\hat\beta$&Bias & SE & SD & Cover   & $\hat\gamma$&Bias & SE & SD & Cover \\ \hline
			1e5 & 20  & 15.55 & 3-stage & 0.806 & 0.76\%  & 0.034 & 0.034 & 94.1\% & 0.197 & -1.49\% & 0.017 & 0.017 & 94.2\% \\
			&              &       & CMLE    & 0.799 & -0.14\% & 0.033 & 0.034 & 94.9\% & 0.201 & 0.28\%  & 0.017 & 0.017 & 95.2\%\\
			&    &                &         &       &        &       &       &       &       &        &       &       &       \\
			1e5 & 50  & 4.51  & 3-stage & 0.818 & 2.21\%  & 0.038 & 0.037 & 91.9\% & 0.197 & -1.62\% & 0.008 & 0.008 & 93.2\% \\
			&               &       & CMLE    & 0.801 & 0.11\% & 0.036 & 0.036 & 95.3\% & 0.200 & 0.02\%  & 0.007 & 0.007 & 94.6\%\\
			\hline
	\end{tabular}}
\end{table}

\begin{table}[!h]\centering
	\caption{Simulation results for estimating the confounding bias parameter $\gamma=0.2$ and the ETT $\beta=0.8$ under Gaussian mixture error based on 1000 replicates. See the caption of Table \ref{table-sim} for description of the summary statistics.}
	\label{table-sim-mixture}
	\begin{tabular}{cccccccccccccccccccr}
		\hline
		$n$& $\eta_z$ & $\hat{\beta}$ &Bias &SE &SD & Cover &$\hat{\gamma}$ &	Bias &	SE &	SD&	Cover\\
		\hline 
		& 0.10   & 0.771 & $-3.64$\% & 0.373 & 0.195 & 75.6\% & 0.256 & 27.78\%  & 0.314 & 0.211 & 66.7\% \\
		1e4  & 0.25    & 0.798 & $-0.29$\% & 0.056 & 0.059 & 93.9\% & 0.205 & 2.39\%  & 0.044 & 0.049 & 91.4\% \\
		& 0.50     & 0.799 & $-0.12$\% & 0.039 & 0.039 & 95.9\% & 0.201 & 0.69\%  & 0.026 & 0.026 & 96.0\% \\
		\midrule
		& 0.10  & 0.780 & $-2.54$\% & 0.106 & 0.084 & 77.0\% & 0.230 & 14.95\%  & 0.090 & 0.101 & 69.2\% \\
		3e4 & 0.25   & 0.799 & $-0.09$\% & 0.032 & 0.033 & 95.1\% & 0.202 & 0.80\%  & 0.025 & 0.027 & 92.9\% \\
		& 0.50    & 0.799 & $-0.07$\%  & 0.023 & 0.022 & 95.4\% & 0.201 & 0.36\% & 0.015 & 0.015 & 95.8\% \\
		\midrule
		& 0.10  & 0.793 & $-0.86$\% & 0.031 & 0.041 & 78.0\% & 0.210 & 4.96\%  & 0.026 & 0.052 & 69.5\% \\
		1e5 & 0.25  & 0.800 &  0.03\% & 0.018 & 0.018 & 94.7\% & 0.200 & 0.24\%  & 0.014 & 0.015 & 93.5\% \\
		& 0.50  & 0.800 &  0.03\%  & 0.012 & 0.012 & 95.3\% & 0.200 & 0.14\% & 0.008 & 0.008 & 95.6\% \\
		\hline
	\end{tabular}
\end{table}


\section{An Application to the Large-Scale UK Biobank Study Data}
UK Biobank is a large-scale ongoing prospective cohort study that recruited around 500,000 participants aged 40-69 in 2006-2010. Participants provided biological samples, completed questionnaires, underwent assessments, and had nurse led interviews. Follow up is chiefly through cohort-wide linkages to National Health Service data, including electronic, coded death certificate, hospital and primary care data \citep{sudlow2015uk}. To control for population stratification, we restricted our analysis to participants with self-reported and genetically validated white British ancestry. For quality control, we also excluded participants with (1) excess relatedness (more than 10 putative third-degree relatives) or (2) mismatched information on sex between genotyping and self-report, or (3) sex-chromosomes not XX or XY, or (4) poor-quality genotyping based on heterozygosity and missing rates $> 2\%$.

To illustrate our methods, we extracted a total of 289 010 white British subjects from the UK Biobank data with complete measurements in body mass index (BMI) and glucose levels. We further selected BMI associated SNP potential IVs among the list provided by \citet{sun2019type}. We found seven SNPs (in Table \ref{table:UKbio}) that are predictive of the residual variance of glucose levels in our Stage 2 regression ( with p-values $P < 0.01$ ), that is, those seven SNPs might satisfy assumption (B3).  The average BMI was 27.39 $kg/m^2$ (SD: 4.75 $kg/m^2$), and the average glucose level was 5.12  $mmol/L$ (SD: 1.21 $mmol/L$). We applied MR MiSTERI to this data  with the goal of evaluating the causal relationship between BMI as treatment and glucose level as outcome; analysis results are summarized in Table \ref{table:UKbio}.

 For comparison purposes, we also include analysis results for  standard two-stage least square (TSLS) implemented in the R package {\it ivreg} \citep{ivreg}. The allele score for the seven SNPs selected is defined as the signed sum of their minor alleles, where the sign is determined by the regression coefficient in our stage 2 regression.  We also implemented MR-GENIUS method \citep{Tchetgen2020_GENIUS} using the same seven SNPs, the causal effect of BMI on glucose was estimated as 0.046 (se: 0.01, 95\% CI: (0.0264,0.0656),  $P$-value: $2.24\times 10^{-6}$). Crude regression analysis by simply regressing glucose on BMI gives an estimate of $0.041$ (se: $4.65\times 10^{-4}$, 95\% CI: (0.0401, 0.0419), $P$-value: $<2\times 10^{-16}$). Further, we included all seven SNPs into a conventional regression together with BMI, and obtained a similar estimate $0.039$ (se: $4.65\times 10^{-4}$, 95\% CI: (0.0381, 0.0399), $P$-value: $<2\times 10^{-16}$). It is interesting to consider results obtained by selecting each SNP as single candidate IV as summarized in the table. For instance, take the SNP rs2176040, the corresponding causal effect estimate is $\hat{\beta} =0.041$ (se: 0.0139, 95\% CI: (0.0138, 0.0682), $P$-value: 0.003), and the selection bias parameter is estimated as $\hat{\gamma}:0.059$ (se: 0.0098, 95\% CI: (0.0398, 0.0782), $P$-value: $2.93\times 10^{-9}$). However, the  TSLS method gives causal effect estimate $-1.619$ (se: 2.04, 95\% CI: (-5.6174, 2.3794), $P$-value: 0.428). These conflicting findings may be due to SNP rs2176040 being an invalid IV, in which case TSLS likely yields biased results. We further combine 20 SNPs, 13 of which are weakly associated with the residual variance of glucose levels, we obtain using CMLE  a causal effect estimate of BMI on glucose  $\hat{\beta}=0.028$ (se: 0.0025, 95\% CI: (0.0231, 0.0329), $P$-value: $6.87\times 10^{-31}$), and a selection bias estimate of $\hat{\gamma} = 0.009$ (se: 0.0017, 95\% CI: (0.0057, 0.0123), $P$-value: $2.82\times 10^{-7}$).  The consistency and asymptotic normality condition  $\hat \kappa=23.72$, greater than our recommended value 10. This final analysis suggests a somewhat smaller treatment effect size than all the other MR methods, while providing strong statistical evidence of a positive effect with a relatively small but statistically significant amount of confounding bias detected.

\begin{table}[h]\centering
	\caption{UK Biobank data analysis results. Columns 2-5 contain the results based on the model (\ref{eq: normal}), columns 6-9 contain results based on the Gaussian mixture model described in Section 2.4, columns 10-11 contain results using the standard two-stage least squares using the R package {\it ivreg}.}
	\label{table:UKbio}
	\begin{tabular}{lcccccccccc}
		\hline
		SNP        & $\hat{\beta}$  & $se(\hat{\beta})$ & $\hat{\gamma}$  & $se(\hat{\gamma})$ & $\hat{\beta}^*$  & $se(\hat{\beta}^*)$ & $\hat{\gamma}^*$  & $se(\hat{\gamma}^*)$  & $\hat{\beta}^{**}$ & $se(\hat{\beta}^{**})$\\
		\hline
		rs3736485  & 0.041 & 0.013 & 0.0001  & 0.0094 & 0.027 & 0.001 & 0.0031  & 0.0002 & 0.200  & 0.068 \\
		rs1558902  & 0.023 & 0.006 & 0.0128  & 0.0046 & 0.040 & 0.001 & 0.0025  & 0.0002 & 0.064  & 0.009 \\
		rs1808579  & 0.041 & 0.014 & 0.0001  & 0.0098 & 0.026 & 0.001 & 0.0042  & 0.0002 & 0.107  & 0.038 \\
		rs13021737 & 0.051 & 0.012 & -0.0065 & 0.0084 & 0.036 & 0.001 & -0.0043 & 0.0005 & 0.030  & 0.016 \\
		rs2176040  & 0.041 & 0.014 & 0.0585  & 0.0099 & 0.026 & 0.001 & 0.0009  & 0.0003 & -1.619 & 2.041 \\
		rs10968576 & 0.047 & 0.016 & -0.0040 & 0.0111 & 0.046 & 0.001 & -0.0005 & 0.0004 & 0.079  & 0.025 \\
		rs10132280 & 0.030 & 0.013 & 0.0081  & 0.0094 & 0.062 & 0.004 & -0.0228 & 0.0005 & 0.049  & 0.029 \\
		Allele score     & 0.030 & 0.003 & 0.0078  & 0.0021 & 0.033 & 0.005 & 0.0105  & 0.0002 & 0.089  & 0.009  \\
		\hline
	\end{tabular}
\end{table}

\section{Discussion} 
In this paper, we have proposed a novel MR MiSTERI method for identification and inference about causal effects in observational studies. MR MiSTERI leverages a possible association between candidate IV SNPs and the variance of the outcome in order to disentangle confounding bias from the causal effect of interest. Importantly, MR MiSTERI can recover unbiased inferences about the causal effect in view even when none of the candidate IV SNPs is a valid IV. Key assumptions which the approach relies on include no additive interaction involving a candidate IV SNP and the treatment causal effect in the outcome model, and a requirement that the amount of selection bias (measured on the odds ratio scale) remains constant as a function of SNP values. Violation of either assumption may result in incorrect inferences about treatment effect. Although not pursued in this paper, robustness to such violation may be assessed via a sensitivity analysis in which the impact of various departures from the assumption may be explored by varying sensitivity parameters. Alternatively, as illustrated in our application, providing inferences using a variety of MR methods each of which relying on a different set of assumptions provides a robustness check of empirical findings relative to underlying assumptions needed for a causal interpretation of results.  We did not include the  MR GxE  method for numerical comparison because it requires users to specify an observed covariate interacting with the SNP IV.  For a particular data analysis problem as we did in Section \ref{sec:UKB}, we typically do not have sufficient prior knowledge  about which  covariate (out of thousands) might interact with a given SNP. In contrast, MR GENIUS method requires that the treatment variable admits a residual which is heteroscedastic with respect to the IV, while MR MiSTERI requires that the outcome admits a residual that is  heteroscedastic with respect to the IV. In practice, one can in principle perform a formal statistical test to determine which method might be most suitable for a particular data analysis problem. Note that both MR MiSTERI and MR GENIUS methods do not require us to specify any covariate a priori interacting with the candidate SNP IV, and hence are more convenient for routine data analyses. 

A notable robustness property of CMLE  estimator of MR MiSTERI is that as we formally establish, it is robust to many weak IV bias, thus providing certain theoretical guarantees of reliable performance in many common MR settings where one might have available a relatively large number of weak candidate IVs, many of which may be subject to pleiotropy.    
While the proposed methods require a correctly specified Gaussian model for the conditional distribution of $Y$ given $A$ and $Z$, such an assumption can be relaxed. In fact, we also consider a Gaussian mixture model as a framework to assess and possibly correct for potential violation of this assumption.  Furthermore, in the Supplementary Materials, we briefly describe a semiparametric three-stage estimation approach under the location-scale model which allows the distribution of standardized residuals to remain unrestricted. It is of interest to investigate robustness and efficiency properties of this more flexible estimator, which we plan to pursue in future work. It is likewise of interest to investigate whether the proposed methods can be extended to the important setting of a binary outcome, which is of common occurrence in epidemiology and other health and social sciences.  

\section*{Acknowledgements}
	Dr. Zhonghua Liu’s research is supported by the Start-up research fund (000250348) of the University of Hong Kong. Dr. Eric Tchetgen Tchetgen is supported by grants R01AG065276, R01CA222147 and R01AI27271. Dr. Baoluo Sun is supported by the National University of Singapore Start-up Grant (R-155-000-203-133). This research has been conducted using the UK Biobank resource (https://www.ukbiobank.ac.uk) under application number 44430.\\
Conflict of interest: None declared. 
\clearpage 

\singlespacing
\bibliographystyle{apalike}
\bibliography{reference}

\newpage
\pagenumbering{arabic}
\setcounter{equation}{0}
\setcounter{table}{0}
\renewcommand{\theequation}{S\arabic{equation}}
\renewcommand{\thetable}{S\arabic{table}}
\begin{center}
	{\large\bf Supplementary Materials}
\end{center}

\section*{Proof of Theorem 1}
\begin{proof}
	Under assumptions (B1) - (B3), we have 
	\begin{equation}
	D(Z=z)= E(Y\mid A=1, Z=z)- E(Y\mid A=0, Z=z) = \beta + \gamma\sigma^2(z). 
	\end{equation}
	Hence, we have the following two equations
	\begin{eqnarray*}
		D(Z=1) &=& \beta + \gamma\sigma^2(Z=1)\\
		D(Z=0) &=& \beta + \gamma\sigma^2(Z=0).
	\end{eqnarray*}
	Solving these two equations simultaneously gives the result in Theorem 1. 
\end{proof}

\section*{Estimation and Inference under Weak Identification} 

Consider the following normal model  also given in the main text
\begin{equation}
\frac{Y- \{ \beta A+ \gamma A \exp(\eta_0+\eta_z Z) + \theta_0 +\theta_z Z\}}{\sqrt{ \exp(\eta_0+\eta_z Z)}} \sim N(0, 1). 
\end{equation}

 Let 
$$ 
I_n(\Theta)=  -E\left\{ \frac{\partial^2}{\partial \Theta\partial \Theta^T} l_n(\Theta)\right\}
$$ 
be the Fisher information matrix for a sample of size $ n $, $ \lambda_{\min} \{I_n(\Theta)\} $ be the minimum eigenvalue of $ I_n(\Theta) $, $tr(A)$ be the trace of a square matrix $A$. 

Condition \ref{cond: mle} summaries the regularity conditions. 
\begin{condition}[Regularity Conditions] \label{cond: mle}
	For every $o= (y, a, z )$ in the support of $O=  (Y, A, Z) $, the likelihood function denoted by $ f(\Theta; o) $ is twice continuously differentiable in $ \Theta $ and satisfies 
	\[
	\frac{\partial }{\partial \Theta} \int \psi_{\Theta} (o) d\nu = \int 	\frac{\partial }{\partial \Theta} \psi_{\Theta} (o) d\nu 
	\]
	for $ \psi_{\Theta} (o)= f_{\Theta} (o) $ and $ = \partial f_{\Theta}(o)/\partial \Theta $, where $ \nu $ is a suitable measure; there is a positive $\epsilon$ such that the Fisher information matrix $ I_n(\Gamma) $ is positive definite for $\Gamma: \|\Gamma- \Theta\|<\epsilon$; and for any given $ \Theta $, there exists a positive number $ \epsilon $ and a positive integrable function $ h $, such that 
	\[
	\sup_{\Gamma: \|\Gamma- \Theta\| < \epsilon} \bigg\| \frac{\partial^2 \log f_{\Gamma} (o)}{\partial \Gamma \partial \Gamma^T} \bigg\| \leq h(o)
	\]
	for all $ o $ in the support of $ O $, where $ \|A\|= \sqrt{tr (A^T A)} $ for any matrix $ A $. 
\end{condition}

Theorem \ref{theo: mle} establishes the consistency and asymptotic normality of solutions to the score function.
\begin{theorem}\label{theo: mle}
	(a) Under Condition \ref{cond: mle} and assume that the observed data $O_i, i=1,\dots, n$ are independent and identically distributed, if  $k/[\lambda_{\min}  \{n I_1(\Theta)\}]\rightarrow0$, 
	then the sequence of conditional maximum likelihood estimators  $\{\hat\Theta_n\}$ is consistent, i.e., $\hat\Theta_n\xrightarrow{p} \Theta.$\\
	(b) Any consistent sequence $\tilde\Theta_n$ such that $S_n(\tilde\Theta)=0$ is asymptotically normal, and as $ n\rightarrow\infty $, 
	\begin{eqnarray}
	\sqrt{n} (\tilde{\Theta}- \Theta) \xrightarrow{d} N(0, \{ I_1 (\Theta)\}^{-1} ). 
	\end{eqnarray}
\end{theorem}
In fact, the general consistency condition in Theorem \ref{theo: mle} can be written as $ \lambda_{\min} \{n I_1(\Theta)\}\rightarrow\infty $ and 
\begin{equation}
\frac{tr\left[I_n^{-1}(\Theta)E\{ S_n(\Theta) S_n(\Theta)^T\}\right] }{\lambda_{\min} \{n I_1(\Theta)\}}\rightarrow 0 \label{seq: general condition}
\end{equation}	
for general likelihood-based methods, which, for example,  applies to the profile-likelihood estimator in \citet{zhao2018statistical}. Specifically, in \citet{zhao2018statistical}, in their notations, $k=1$, $E\{ S_n(\Theta) S_n(\Theta)^T\}= V_1$, $I_n(\Theta)= V_2$, and $V_1=V_2+O(p), V_2\asymp n \|\gamma\|^2$, and thus, condition \eqref{seq: general condition}  becomes $ V_1/V_2^2= O\{p/(n^2\|\gamma\|^4)\} \rightarrow 0$, which is the condition for consistency in \citet[Theorem 3.1]{zhao2018statistical}.

\begin{proof} First, notice that the condition $k/[\lambda_{\min}  \{n I_1(\Theta)\}]\rightarrow0$ implies $\lambda_{\min}  \{n I_1(\Theta)\}\rightarrow\infty$. 
	Let $B_n(c)= \{\Gamma: \|[I_n(\Theta)]^{1/2} (\Gamma- \Theta)\|\leq c \}$ for $c>0$, where $ \| A\|= \sqrt{tr (A^T A)} $ for any matrix $ A $. Since $B_n(c)$ shrinks to $\Theta$ as $n\rightarrow \infty$, the consistency result in Theorem \ref{theo: mle}(a) is implied by the fact that for any $ \epsilon >0$, there exists $ c>0 $ and $ n_0>1 $ such that 
	\begin{align}
	P(l_n(\Gamma)- l_n(\Theta)<0 ~  \text{for all }\Gamma \in \partial B_n(c)) \geq 1-\epsilon, \qquad n\geq n_0 \label{seq: mle}
	\end{align}
	and $\hat\Theta_n$ is unique for $n\geq n_0$, 	where $  \partial B_n(c) $ is the boundary of $ B_n(c) $. For $ \Gamma \in  \partial B_n(c) $, the Taylor expansion gives 
	\begin{align*}
	l_n(\Gamma)- l_n (\Theta) &= c\lambda^T [I_n(\Theta)]^{-1/2} S_n(\Theta) \\
	&+ (c^2/2) \lambda^T  [I_n(\Theta)]^{-1/2}  i_n(\Gamma^*) [I_n(\Theta)]^{-1/2} \lambda,
	\end{align*}
	where $ \lambda= [I_n(\Theta)] ^{1/2} (\Gamma- \Theta)/c$ satisfying $\| \lambda \|= 1$, $ i_n(\Gamma)= \partial^2l_n(\Gamma)/ \partial \Gamma\partial \Gamma^T$, and $ \Gamma^* $ lies between $ \Gamma $ and $ \Theta $. Note that 
	\begin{align}
	E \frac{\|i_n (\Gamma^*)- i_n (\Theta)\|}{n} &\leq E\max_{\Gamma\in B_n(c)} \frac{\|  i_n (\Gamma)- i_n (\Theta) \|}{n}\nonumber \\
	&	\leq  E\max_{\Gamma\in B_n(c)} \| i_1(\Gamma)- i_1(\Theta)\| \rightarrow 0 \label{seq: mle3}
	\end{align}
	which follows from the dominated convergence theorem combined with the facts that (a) $ i_1(\Gamma)= \partial l_1(\Gamma)  /\partial \Gamma \partial  \Gamma^T$ is continuous in a neighborhood of $ \Theta $ for any fixed observation; (b) $ B_n(c) $ shrinks to $ \{\Theta\} $ from $ \lambda_{\min} (I_n(\Theta)) \rightarrow\infty$; and (c) for sufficiently large $ n $, $ \max_{\Gamma \in B_n(c)} \| i_1(\Gamma)- i_1(\Theta) \| $  is bounded by an integrate function under Condition \ref{cond: mle}. By the strong law of large numbers, $ \|n^{-1}  i_n(\Theta) + I_1(\Theta)\| \xrightarrow{a.s.} 0$. These results imply that 
	\begin{align}
	l_n(\Gamma)- l_n (\Theta)= c\lambda^T [I_n(\Theta)]^{-1/2} S_n(\Theta) - (1+o_p(1)) c^2/2. \label{seq: mle2}
	\end{align}
	Note that $ \max_{\lambda}\{ \lambda^T [I_n(\Theta)]^{-1/2}S_n(\Theta)\} = \| I_n(\Theta)^{- 1/2} S_n(\Theta) \|$ with 
	
	$$ \lambda= I_n(\Theta)^{-1/2} S_n(\Theta	) /  \| I_n(\Theta)^{- 1/2} S_n(\Theta) \|.$$
	Therefore, \eqref{seq: mle}  follows from \eqref{seq: mle2} and 
	\begin{align*}
	&  P (   c\lambda^T [I_n(\Theta)]^{-1/2} S_n(\Theta) - c^2/2<0~  \text{for all }\lambda ~ s.t.~ \|\lambda\|=1)\\
	= &  P \left( \max_{\lambda}  \lambda^T [I_n(\Theta)]^{-1/2} S_n(\Theta) <  c/2\right)\\
	=&  P \left(  \| I_n(\Theta)^{- 1/2} S_n(\Theta) \| <  c/2\right)\\
	=& 1- P \left(  \| I_n(\Theta)^{- 1/2} S_n(\Theta) \| \geq   c/2\right)\\
	\geq &  1- \frac{4 E  \| I_n(\Theta)^{- 1/2} S_n(\Theta) \|^2 }{c^2/4} \qquad \text{ (from the Markov inequality) }\\
	=& 1-  \frac{E ~ tr \{ S_n(\Theta)^TI_n(\Theta)^{-1} S_n(\Theta)\}  }{c^2/4}\\
	=&  1-  \frac{ tr \{ I_n(\Theta)^{-1} E[S_n(\Theta) S_n^T(\Theta)]\}  }{c^2/4} \\
	\geq &1-\epsilon,
	\end{align*}
	from choosing $ c $ such that $  c^2/tr \{ I_n(\Theta)^{-1} E[S_n(\Theta) S_n^T(\Theta)]\} $ is large enough, which is possible because 
	\begin{align*}
	&c^2/tr \{ I_n(\Theta)^{-1} E[S_n(\Theta) S_n^T(\Theta)]\}\\
	\geq &  \|[I_n(\Theta)]^{1/2} (\Gamma- \Theta)\|^2 /   tr \{ I_n(\Theta)^{-1} E[S_n(\Theta) S_n^T(\Theta)]\}\\
	\geq &\lambda_{\min} [ I_n(\Theta)] \| \Gamma- \Theta\|^2/  tr \{ I_n(\Theta)^{-1} E[S_n(\Theta) S_n^T(\Theta)]\} \\
	=& \| \Gamma- \Theta\|^2/o(1).
	\end{align*}
	For $n\geq n_0$, $\hat\Theta_n$ is unique because from Condition 2, the Fisher information $I_n(\Gamma)$ is positive definite in a neighborhood of $\Theta$.
	
	(b)  Using the mean value theorem for vector-valued functions, we obtain that 
	\[
	- S_n(\Theta) = \left[ \int_{0}^1  i_n\big(\Theta+ t(\tilde{\Theta} - \Theta)\big)dt\right](\hat{\Theta} - \Theta). 
	\]
	Note that using a similar argument as in \eqref{seq: mle3}, when $ \hat{\Theta} - \Theta =o_p(1) $,
	\[
	\frac{1}{n} \left\|  i_n\big(\Theta+ t(\tilde{\Theta} - \Theta)\big) -  i_n(\Theta)\right\|\xrightarrow{p} 0. 
	\]
	Since $ n^{-1} i_n (\Theta) \xrightarrow{p} - I_1(\Theta)$ and $ I_n (\Theta) = n I_1(\Theta)$, 
	\[
	- S_n(\Theta) = - I_n (\Theta) (\tilde{\Theta} - \Theta) + o_p(  \|  I_n (\Theta) (\tilde{\Theta} - \Theta)\|). 
	\]
	This and Slutsky's theorem imply that $ \sqrt{n } (\tilde{\Theta} - \Theta) $ is asymptotically equivalent with 
	\[
	\sqrt{n} [ I_n (\Theta)]^{-1} S_n(\Theta) = n^{-1/2} [I_1(\Theta)]^{-1} S_n(\Theta) \xrightarrow{d} N_k (0, I[_1(\Theta)]^{-1})
	\]
	by the central limit theorem for i.i.d. samples. 
\end{proof}

\section*{Empirical guidelines for asymptotics} 
In practice, it is important to know what value of $\hat\kappa$ would be considered large enough for the CMLE to perform well, i.e., to have small bias and adequate coverage probability, in presence of many weak invalid IVs. 

\begin{figure}[ht]
    \centering
    \includegraphics[scale=0.8]{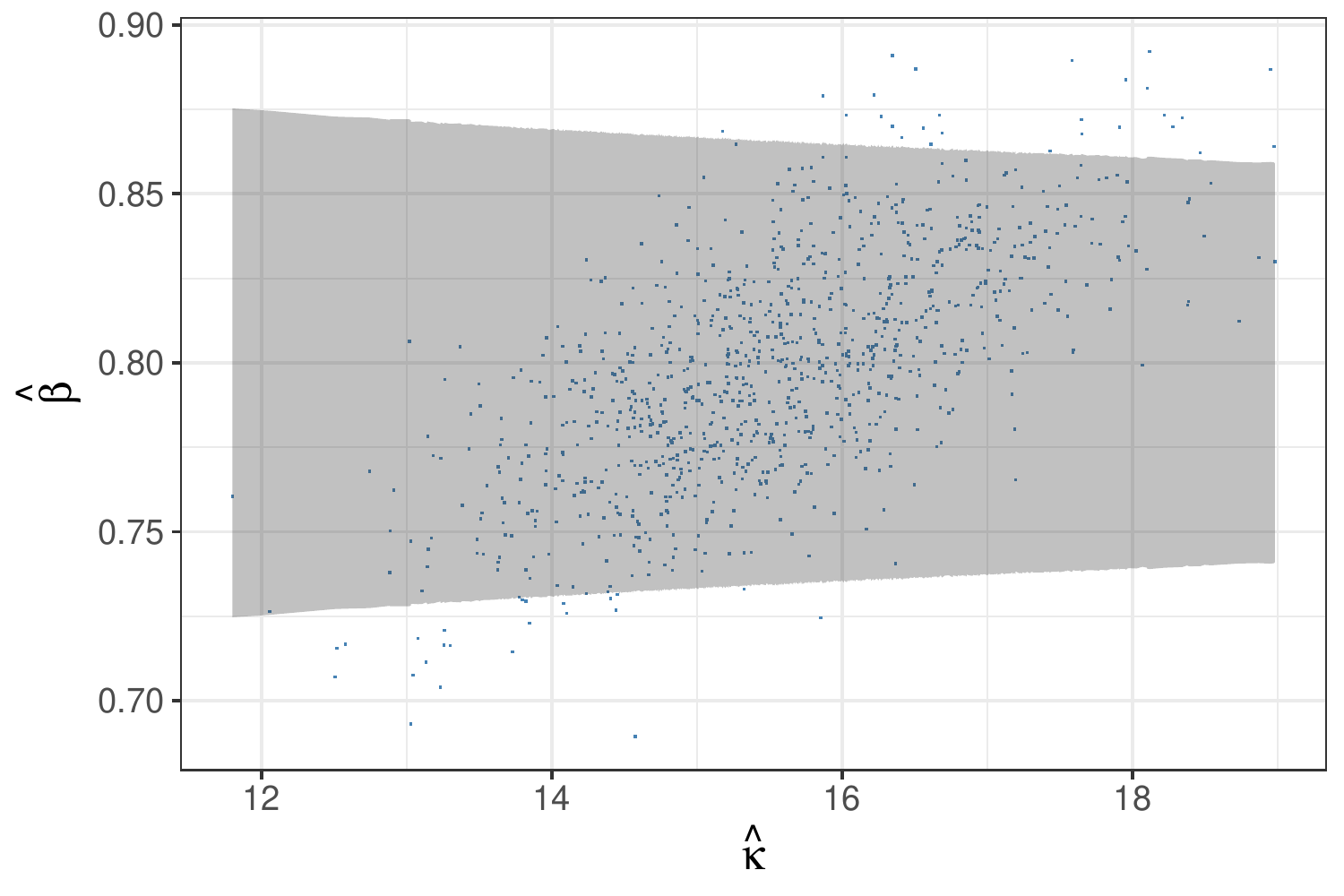}
    \caption{Evaluation of the consistency and asymptotic normality condition for the CMLE under the same setting as the second simulation study in Section 3 of the main article, with $n=10^5, p=20$. The x-axis plots the condition $\hat\kappa$, and the y-axis plots values of the CMLE in 1000 simulations. The shaded area represents the two-sided error bands centered at $\beta=0.8$.     \label{fig:n1e5p20}}
\end{figure}

\begin{figure}[h]
    \centering
    \includegraphics[scale=0.8]{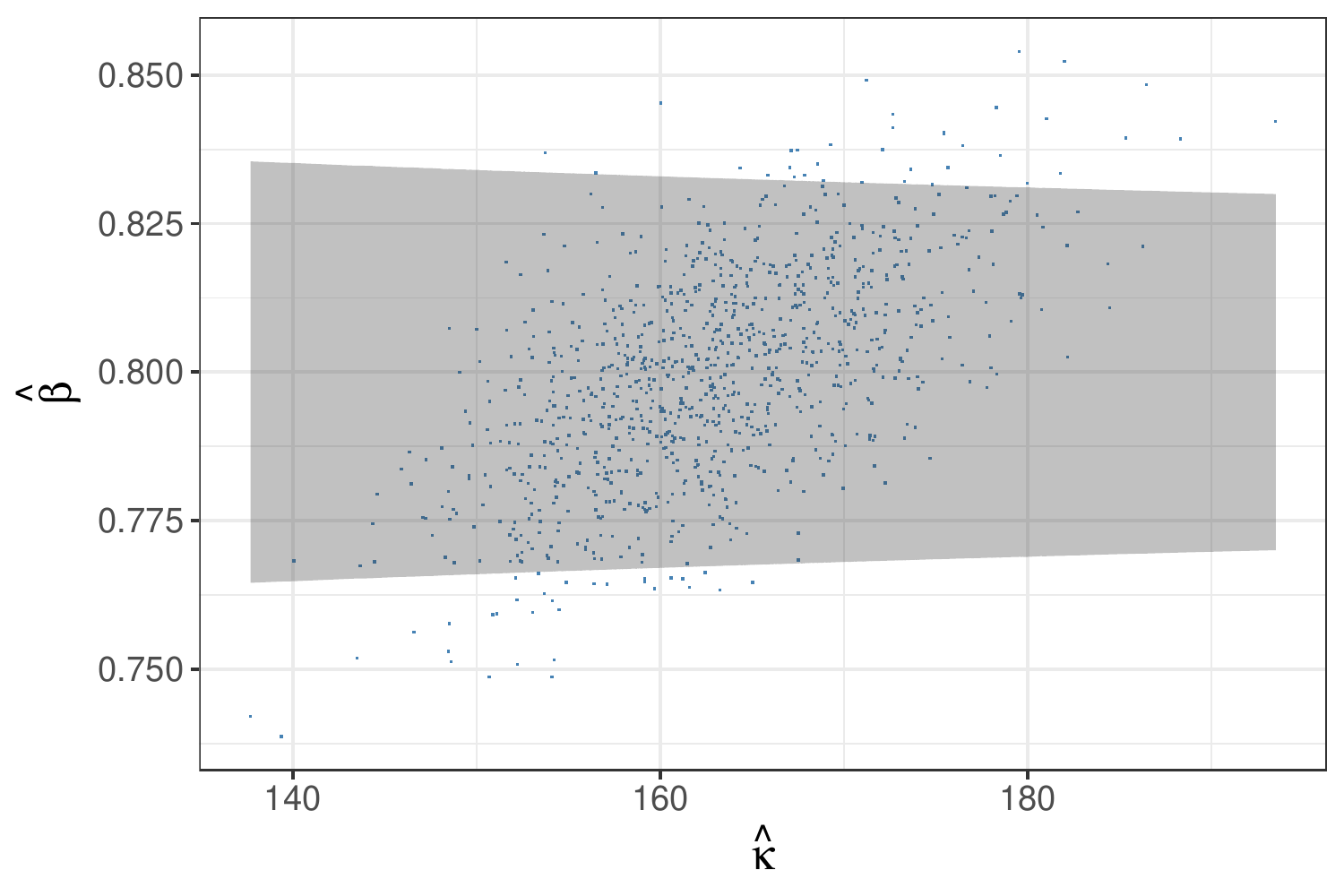}
    \caption{Evaluation of the consistency and asymptotic normality condition for the CMLE under the same setting as the second simulation study in Section 3 of the main article, with $n=10^6, p=5$. The x-axis plots the condition $\hat\kappa$, and the y-axis plots values of the CMLE in 1000 simulations. The shaded area represents the two-sided error bands centered at $\beta=0.8$.     \label{fig:resp5n1e+06eta_z0.05}}
\end{figure}

We follow the same setting as the second simulation study in Section 3 of the main article. We consider two scenarios: (i) $n=10^5, p=20$ in Figure \ref{fig:n1e5p20}; and (ii) $n=10^6, p=5$ in Figure \ref{fig:resp5n1e+06eta_z0.05}. For each of the 1000 simulation runs, we plot $\hat\beta$ against $\hat\kappa$. We also plot the two standard error bands centered at $\beta_0$ (shaded area). From Figures \ref{fig:n1e5p20}-\ref{fig:resp5n1e+06eta_z0.05}, $\hat\kappa$ are all larger than 10, and $\hat\beta$ are inside the shaded area, close to the true value $\beta=0.8$. Note that because $\hat\kappa$ (which relates to the Hessian matrix) is a function of $\hat\beta$, we see that $\hat\beta$ is not symmetric around the true value $\beta=0.8$. But this usually does not affect using $\hat\kappa$ as a condition to evaluate whether the CMLE would perform well. We have also simulated extensively under other settings,
we find that the CMLE has good performance when $\hat\kappa\geq 10$.


\section*{Estimation method under Gaussian mixture error} 
We describe the alternating optimization procedure with user-specified $K$ and parametric models $\sigma(Z=z;\eta)$ and $E(Y|A=0,Z=z;\theta)\equiv\mu_0(z;\theta)$  indexed by finite-dimensional parameters $\eta$ and $\theta$ respectively. 

\begin{itemize}
	\item[(i)]Initialization step: maximize the
	standard Gaussian location scale model 
	\[
	\frac{Y-\{\beta A+\gamma A\sigma ^{2}\left( Z;\eta\right) +\mu_0(Z;\theta)\} }{\sigma \left( Z;\eta\right) }\sim N\left( 0,1\right) 
	\]
	with respect to $(\beta,\theta,\eta,\gamma)$ by maximum likelihood estimation, to obtain $(\tilde{\beta},\tilde{\theta},\tilde{\eta},\tilde{\gamma})$.  Construct the standardized residuals $$\tilde{\epsilon}=\frac{Y-\{\tilde{\beta} A+\tilde{\gamma} A\sigma ^{2}\left( Z;\tilde{\eta}\right)+\mu_0(Z;\tilde{\theta})\} }{\sigma\left( Z;\tilde{\eta}\right)}.$$
	\item[(ii)] Maximize the Gaussian mixture location scale model
	\[
	\tilde{\epsilon}\sim \sum_{k=1}^K \pi_k N\left( \mu_k,\delta_k^2\right) 
	\]
	with respect to $(\pi_1,...,\pi_K,\mu_1,...,\mu_K,\delta_1,...,\delta_K)$ using  constrained maximum likelihood estimation to obtain $(\tilde{\pi}_1,...,\tilde{\pi}_K,\tilde{\mu}_1,...,\tilde{\mu}_K,\tilde{\delta}_1,...,\tilde{\delta}_K)$. Construct 
	\begin{equation*}
	\begin{aligned}
	\tilde{v} =A\sum_{k=1}^{K}\frac{\tilde{\pi}_{k}\tilde{\omega}_{k}\tilde{\delta}_{k}^{2}\sigma
		^{2}\left( Z;\tilde{\eta}\right) }{\sum_{k=1}^{K}\tilde{\pi}_{k}\tilde{\omega}_{k}},\quad \tilde{w} =\frac{\sigma \left(
		Z;\tilde{\eta}\right) }{\sum_{k=1}^{K}\tilde{\pi}_{k}\tilde{\omega}_{k}}\left\{ \left[
	\sum_{k=1}^{K}\tilde{\pi}_{k}\tilde{\omega}_{k}\tilde{\mu} _{k}\right] \right\},
	\end{aligned}
	\end{equation*}
	where $$\tilde{\omega}_k =\exp \{ \tilde{\gamma} A\sigma \left( Z;\tilde{\eta}\right) \tilde{\mu}_{k} + \tilde{\delta}_{k}^{2}\left[ \tilde{\gamma} A \sigma \left( Z;\tilde{\eta}\right) \right] ^{2}/2\}.$$
	
	\item[(iii)] Minimize $n^{-1}\sum_i\{Y_i-\beta A_i-\mu_0(Z_i;\theta)-\gamma \tilde{v}_i-\tilde{w}_i\}^2$ using the least squares method with offset $\tilde{w}$ to obtain new $(\tilde{\beta},\tilde{\theta},\tilde{\gamma})$, followed by regressing the squared residuals to obtain new $\tilde{\eta}$.  Construct the standardized residuals $$\tilde{\epsilon}=\frac{Y-\{\tilde{\beta} A+\mu_0(Z;\tilde{\theta})+\tilde{\gamma} \tilde{v}+\tilde{w}\}}{\sigma\left( Z;\tilde{\eta}\right)},$$
	based on the new values $(\tilde{\beta},\tilde{\theta},\tilde{\gamma}, \tilde{\eta})$.
	
	\item[(iv)] Iterate between steps (ii) and (iii) until change in log-likelihood derived at step (ii) falls below a user-specified tolerance level. In this simulation we iterate until $|\log LL_j-\log LL_{j-1}|/|\log LL_{j-1}|<0.001$ at the $j$-th iteration.
\end{itemize}
Asymptotic variance is estimated based on standard M-estimation theory by stacking the estimating functions in steps (ii) and (iii), evaluated at the final parameter values at convergence.

\section*{Semiparametric three-stage estimation} 

Consider the general location-scale model 
$$Y=E(Y|A,Z)+\sigma(Z)\epsilon\equiv \mu(A,Z)+\sigma(Z)\epsilon,$$ where $\epsilon \perp A,Z$. Under assumptions (B1)-(B3), the conditional mean function is given by
\begin{eqnarray*}
	\mu(a,z)&=&\beta a+\mu(0,z)
	+\frac{\partial}{\partial (\gamma a)}\ln E \{\exp \left( \gamma a\sigma \left(
	z\right) \epsilon \right)|A=0,Z=z\} \\
	&=&\beta a+\mu(0,z)
	+\sigma \left( z\right) \frac{E\left\{ \epsilon \exp \left( \gamma
		a\sigma \left( z\right) \epsilon \right) \right\} }{E\left\{ \exp \left(
		\gamma a\sigma \left(z\right) \epsilon \right) \right\} }.
\end{eqnarray*}%
A semiparametric three-stage estimator of $(\beta,\gamma)$ may be implemented via the following steps:

\begin{itemize}
	\item[(i)] Fit the regression model $\mu(a,z)=m_{a}\left( a\right)
	+m_{z}\left( z\right) +m_{az}\left( a,z\right) $, where $0=m_{z}\left(
	0\right) =m_{az}\left( a,0\right) =m_{az}\left(0,z\right) $ for
	identification, using a nonparametric method such as generalized additive model. For example, if the support of $Z$ is $\{0,1,2\}$, then the nonparametric model for the outcome is of the form $\mu(a,z) =\ m_{a}\left( a\right)+\{\gamma_1 z+\gamma_2 z^{2}\} +\{z m_{1}\left( a\right) +z^{2}m_{2}\left(
	a\right)\} $ with $m_{1}\left( 0\right) =m_{2}\left( 0\right) =0;$ a saturated model may be considered if $A$ also has finite support. Let $\widehat{\mu }\left(a,z\right) $ denote the resulting
	estimator of the mean  $\mu(a,z)$.
	\item[(ii)] Using the residuals from (i), fit a nonparametric model for the conditional variance $\sigma
	^{2}\left( z\right) $. If the support of $Z$ is $\{0,1,2\}$, then we can specify the saturated model $\sigma^2(z)=\exp(\eta_0+\eta_1 z +\eta_2 z^2)$. Let $\widehat{\sigma }^{2}\left(
	z\right)$ denote the resulting estimator of $\sigma^2(z)$. 
	\item[(iii)] Define $\widehat{\epsilon }_{j}=\left\{ Y_{j}-\widehat{\mu }\left(
	A_{j},Z_{j}\right) \right\} /\widehat{\sigma }\left(Z_{j}\right) $ and
	let 
	\[
	\widehat{g}_{i}\left( \gamma \right) =\frac{\sum_{j=1}^{n}\widehat{\epsilon }%
		_{j}\exp \left( \gamma A_{i}\widehat{\sigma }\left( Z_{i}\right) 
		\widehat{\epsilon _{j}}\right) }{\sum_{j=1}^{n}\exp \left( \gamma A_{i}%
		\widehat{\sigma }\left( Z_{i}\right) \widehat{\epsilon }_{j}\right) },\quad i=1,...,n.
	\]
	The proposed semiparametric three-stage estimator of $(\beta,\gamma)$ is
	\[
	(\hat{\beta},\hat{\gamma})=\arg \min_{\beta , \gamma }\sum_{i}\left\{
	Y_{i}-\beta A_{i}-\widehat{\mu}(0,Z_i)-\widehat{\sigma }\left(
	Z_{i}\right) \widehat{g}_{i}\left( \gamma \right) \right\} ^{2}.
	\]%
\end{itemize}

\noindent We note that a potentially more efficient approach is to maximize the log-likelihood for the model using a kernel estimator for the density of standardized residuals from step (iii). Specifically, let $$\widehat{\delta}_{i}(\beta,\gamma)=\{ Y_{j}-\beta A_{i}-\widehat{\mu}(0,Z_i)-\widehat{\sigma }\left(
Z_{i}\right) \widehat{g}_{i}\left( \gamma \right) \} /\widehat{\sigma }\left(Z_{i}\right) $$ and define $$\ell_n(\beta,\gamma) =\sum_{i=1}^n \log \widehat{f}_h(\widehat{\delta}_{i}(\beta,\gamma)),$$
where $\widehat{f}_h(\epsilon)=\frac{1}{nh}\sum_{i=1}^n K\left(\frac{\epsilon-\widehat{\epsilon }_{i}}{h}\right)$ is a kernel density estimator with bandwidth $h>0$. The alternative estimator of $(\beta,\gamma)$ is given by
\[
(\hat{\beta},\hat{\gamma})=\arg \max_{\beta , \gamma }\left\{\ell_n(\beta,\gamma)\right\}.
\]%

\end{document}